\documentclass{ws-jai}
\usepackage[flushleft]{threeparttable}
\begin{document}

\catchline{}{}{}{}{} % Publisher's Area please ignore

\markboth{Author's Name}{Paper Title}

\title{Multi-Level Pre-Correlation RFI Flagging for Real-Time Implementation on UniBoard}

\author{C\'{e}dric Dumez-Viou$^\dagger$, Rodolphe Weber$^{\ddagger,\dagger}$, Philippe Ravier$^\ddagger$}

\address{
$^\dagger$Observatoire de Paris, Station de radioastronomie de Nan\c{c}ay, 18330 Nan\c{c}ay, France\\
$^\ddagger$PRISME laboratory,University of Orl\'{e}ans, 12 rue de Blois BP 6744, Orl\'{e}ans, France
}

\maketitle

\corres{$^\dagger$Corresponding author.}

\begin{history}
\received{(to be inserted by publisher)};
\revised{(to be inserted by publisher)};
\accepted{(to be inserted by publisher)};
\end{history}

\begin{abstract}
Because of the denser active use of the spectrum, and because of radio telescopes higher sensitivity, radio frequency interference (RFI) mitigation has become a sensitive topic for current and future radio telescope designs. Even if quite sophisticated approaches have been proposed in the recent years, the majority of RFI mitigation operational procedures are based on post-correlation corrupted data flagging. Moreover, given the huge amount of data delivered by current and next generation radio telescopes, all these RFI detection procedures have to be at least automatic and, if possible, real-time.

In this paper, the implementation of a real-time pre-correlation RFI detection and flagging procedure into generic high-performance computing platforms based on Field Programmable Gate Arrays (FPGA) is described, simulated and tested. One of these boards, UniBoard,  developed under a Joint Research Activity in the RadioNet FP7  European programme is based on eight FPGAs interconnected by a high speed transceiver mesh. It provides up to ~4 TMACs with \textregistered{Altera} Stratix IV FPGA and 160 Gbps data rate for the input data stream.

The proposed concept is to continuously monitor the data quality at different stages in the digital preprocessing pipeline between the antennas and the correlator, at station level and core level.  In this way, the detectors are applied at stages where different time-frequency resolutions can be achieved and where the interference-to-noise ratio is maximum, right before any dilution of RFI characteristics by subsequent channelizations or signal recombinations. The detection decisions could be linked to a RFI statistics database or could be attached to the data for later stage flagging.

Considering the high in-out data rate in the pre-correlation stages, only real-time and go-through detectors (\textit{i.e.} no iterative processing) can be implemented. In this paper, a real-time and adaptive detection scheme is described. 

An ongoing case study has been set up with the Electronic Multi-Beam Radio Astronomy Concept (EMBRACE) radio telescope facility at Nan\c{c}ay Observatory. The objective is to evaluate the performances of this concept in term of hardware complexity, detection efficiency and additional RFI metadata rate cost. The UniBoard implementation scheme is described.

\end{abstract}


\section{Introduction}
The sensitivity of state-of-the-art telescopes is over ten orders of magnitude higher than in most communications systems. Although radio telescopes are best located in relatively remote areas, astronomical observations may still be hampered
by man-made radio frequency interference (RFI). Under this assumption, one basic or recurrent scenario could be to carefully design the analogue parts and  to limit the full-time digital measures to flagging. 

In particular, it would be worthwhile to continuously monitor the quality of the data. Given the extreme sensitivity of current and future radio telescopes, this task has to be a by product of the instrument itself. It would therefore be interesting to implement some detection methods as regular signal processing tasks at station or core levels. The detection results could be linked to a kind of RFI statistics database or attached to the data for flagging.

An exhaustive survey on RFI mitigation for radio astronomy \cite{boonstra_2009_1} and a workshop \cite{rfi_2010} dedicated to that topic have shown various developments for preserving radio astronomical observations.

Currently, in routine practice, the majority of RFI mitigation procedures are based on corrupted data detection and flagging at post correlation level. For example, the Low Frequency Array (LOFAR) provides a powerful post-correlation RFI classification tool based on combinatorial thresholding \cite{offringa_2010_1}.

\begin{figure}[h]
\begin{center}
\includegraphics[width=\linewidth]{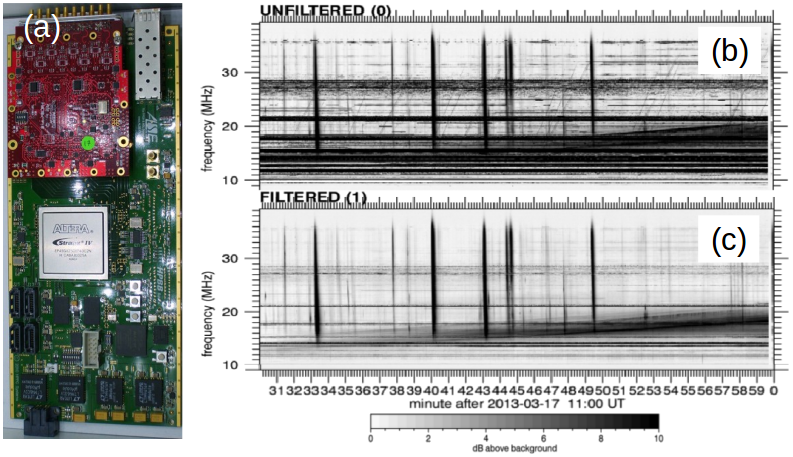} %100 percent
\end{center}
\caption{Spectrum Observation with the Nan\c{c}ay decameter array radio telescope with real time post correlation RFI mitigation. (a) The FPGA board, based on \textregistered{Altera} Stratix IV GX,  has been developed by ALSE \cite{alse_2016}. The ADC daughter board provides 14 bits samples at 80~MHz. The FPGA processes in real-time a 320~MB/s stream to compute 65536-bins power spectra filtered by two stages of median filtering (frequency domain and then, time domain) to provide robust data reduction (b) Time-frequency plane without median filtering. (c) Time-frequency plane with median filtering.}
\label{fig:DAM_median}
\end{figure}

Generally, post-correlation processing approaches are based on iterative algorithms including sorting and hierarchical/scaling strategies \cite{offringa_2010_2,offringa_2012_1, floer_2010_1, fridman_2010_1}. Figure~\ref{fig:DAM_median} shows our implementation into a FPGA board of a median filter on high-resolution power spectrum observations at the Nan\c{c}ay decameter array radio telescope. However, the high data-rate at the pre-correlation stages does not allow efficient implementations of these approaches due to their computational complexities.

In phased array radio telescopes, pre-correlation samples are labelled with parameters such as time, frequency, polarisation, steering direction, antenna or station number.  For example, phased array-based interferometers, these samples are called beamlets. No continuity in frequency or steering direction is guaranteed between beamlets (\textit{i.e.} two consecutive beamlets can represent completely different frequencies and steering directions).

\begin{figure}[h]
\begin{center}
\includegraphics[width=0.5\linewidth]{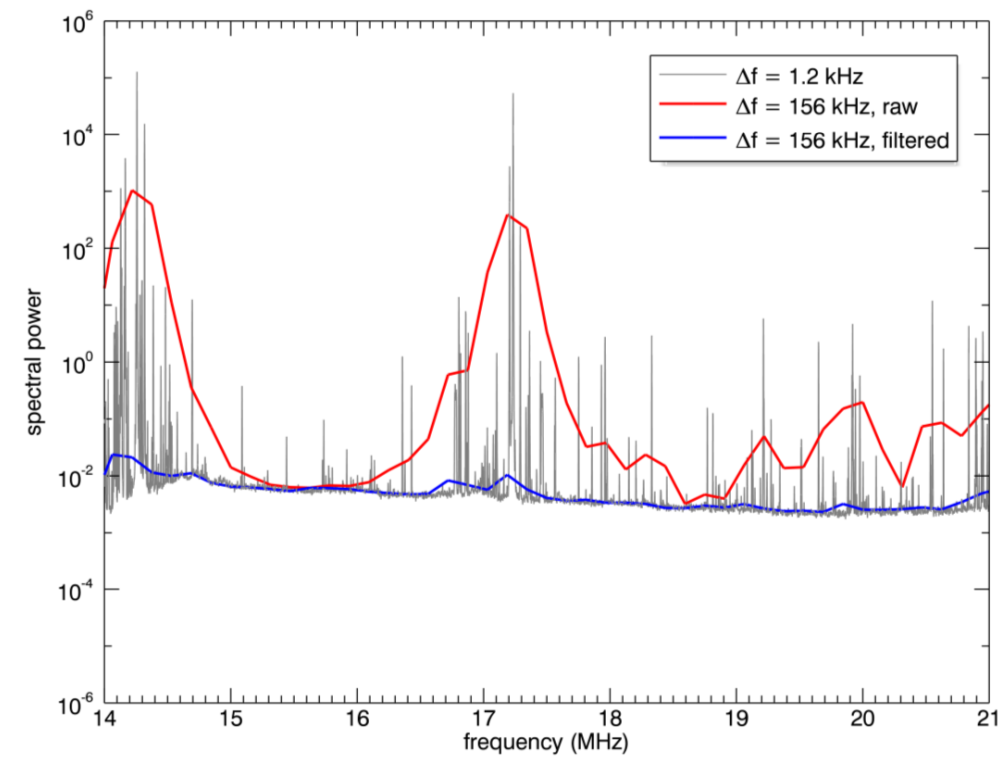} %100 percent
\end{center}
\caption{Spectrum Observation with the Nan\c{c}ay decameter array radio telescope. RFI can be precisely discriminated at high frequency resolution ($\Delta f=1.2$~kHz, gray line) to provide a recombined filtered spectrum ($\Delta f=156$~kHz, blue line). Direct RFI mitigation with $\Delta f=156$~kHz (red line) will blank most of the observation band or will miss some RFI peaks.}
\label{fig:freq_resolution}
\end{figure}

\begin{figure}[h]
\begin{center}
\includegraphics[width=0.75\linewidth]{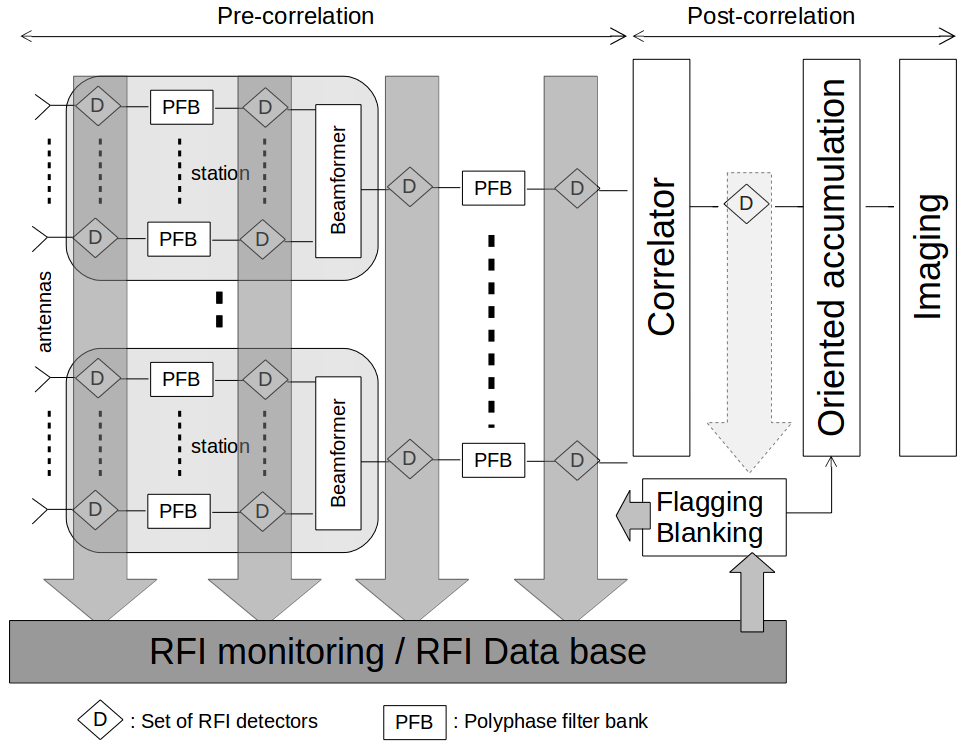} %100 percent
\end{center}
\caption{Ideal overview of the Multi-level Pre-correlation RFI monitoring concept. At each stage, a set of RFI detectors monitor the data flow in real-time. RFI flags are processed to control the correlator integration behaviour (\emph{oriented accumulation}). Optionally, RFI statistics are sent to a data base for further off-line analysis or used, straight, for blanking.}
\label{fig:overview}
\end{figure}

In the proposed approach, each beamlet will be, independently and simultaneously, processed at different data processing stages. Therefore, the hierarchical/scaling strategy will still be possible from the antenna up to the correlator through all of the filter banks and beamformers which are implemented in between. The different detectors can be activated or not according to the RFI context of the observation. Thus, local RFI will be best detected at affected remote stations before correlating their signals with other stations, while broadband RFI will be better detected at an early stage in the channelisation process, and so on...  In this way, the detectors are applied at stages where different time-frequency-space RFI parameters can be naturally achieved and where the interference-to-noise ratio is maximum (see Figure~\ref{fig:freq_resolution}), right before any dilution of RFI characteristics by subsequent channelizations or signal recombinations. Figure~\ref{fig:overview} gives an overview of the proposed approach.

These detection flags can be used in different ways:
\begin{itemize}
\item as metadata uploaded in a specific RFI database. The latter could be used to  analyse, \textit{a posteriori}, the data quality and consequently the observation quality.
\item as flags for local RFI mitigation, such as triggering some spatial nulling a station level.
\item as a global flag which will be propagated up to the correlator stage. These global flags would be used to selectively integrate the corresponding data (\emph{oriented accumulation}). Then, several sets of visibilities or covariance matrices with various expected quality levels will be provided and no data will be lost.  According to the type of observations, these datasets could be merged or not for further processing (calibration, imaging).
\end{itemize}

\begin{figure}[h]
\begin{center}
\includegraphics[width=\linewidth]{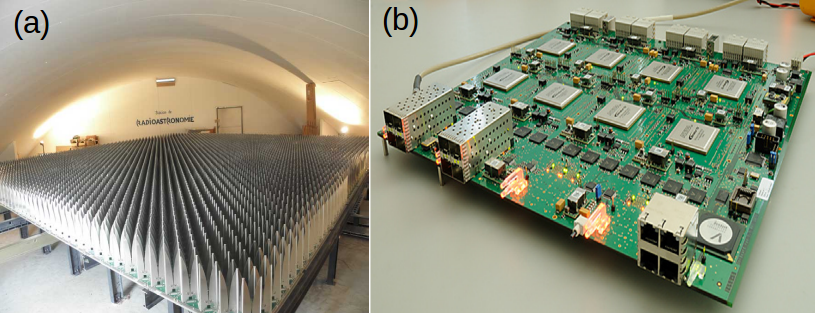} %100 percent
\end{center}
\caption{\textbf{(a)} The French EMBRACE demonstrator.  \textbf{(b)} The UniBoard. }
\label{fig:embrace_uniboard}
\end{figure}

To experiment with all these concepts, an ongoing case study has been conducted on the Electronic Multi-Beam Radio Astronomy Concept (EMBRACE) radio telescope facility at Nan\c{c}ay Observatory (see Figure~\ref{fig:embrace_uniboard}.a and \cite{torchinsky_2016_1}). The objective of the experiment is to evaluate the RFI mitigation strategy in terms of hardware complexity, detection efficiency and additional RFI metadata datarate cost.

All the pre-correlation processing is implemented into the UniBoard (see Figure~\ref{fig:embrace_uniboard}.b and \cite{szomoru_2010_1,hargreaves_2012_1}) , a generic high-performance computing platform based on Field Programmable Gate Arrays (FPGA). This board, developed under a Joint Research Activity in the RadioNet FP7  European programme is based on eight FPGA interconnected by a high speed transceiver mesh. It provides up to 4~TMACs, with \textregistered{Altera} Stratix IV FPGA, and 160~Gbps data rate for the input data stream.

In Section~\ref{theo_descrip} the principles of the pre-correlation power detection is presented. In order to validate the concept, some simulations and a first implementation are also described. In section~\ref{firmware}, the detailed implementation of the EMBRACE-UniBoard experiment is described. Finally, conclusions and perspectives are given in section~\ref{conclusion}.

\section{RFI detection criterion theoretical description} \label{theo_descrip}
 Let us define, $x_k(t)$, the digital signal at a given stage in the radio telescope data stream, $k$ is the index representing the set of parameters (\textit{e.g.} channel frequency, beam number, antenna number, station number, polarisation type...). The problem is formulated as a classical binary hypothesis test where the two simple hypothesis '$H0$ : the Signal-of-Interest (SOI) alone' and '$H1$ : SOI plus RFI' are confronted:

\begin{eqnarray}
H0 : x_{k}(t)=\sigma_n .n(t)&\\
H1 : x_{k}(t)=\sigma_n .n(t)&+\sigma_r . r(t)
\end{eqnarray}

where  $n(t)$ is a independent and identically distributed. normalized (\textit{i.e}. centred and with unit variance) Gaussian complex valued random process and $r(t)$ is a i.i.d unknown complex valued  RFI with unit power. The ratio $\frac{\sigma_r^2}{\sigma_n^2}$ defines the interference-to-noise ratio (INR). We assume herein that $n(t)$ and $r(t)$ are independent.

One way to solve this problem is to use the classical approach pioneered by Neyman and Pearson which maximizes the detection probability ($P_d$=Proba(reject $H0 |H1$ is true) while keeping the false alarm probability ($P_{fa}$=Proba(reject $H0|H0$ is true) fixed at a constant, preferably low,  value $\alpha$.

In general, the optimal solution can be reduced to the comparison of a criterion $C$ to a constant threshold $\eta$. Under Gaussian hypothesis, the criterion $C$ would be:

\begin{equation}
C=\sum_{t=1}^{N} \left| x_k(t) \right|^2 \underset{H_0}{\overset{H_1}{\gtrless}} \eta 
\end{equation}
with the threshold, $\eta$, defined by 
\begin{equation}
\eta = 
\frac{\sigma_n^2}{2}  Q^{-1}_{\chi^2_{2N}}\left( 1-\alpha\right) = \lambda(\alpha) .\sigma_n^2
\label{eq:eta_th}
\end{equation}
where $ Q_{\chi^2_{2N}}$ represents the tail probability of a $\chi^2$ distribution with $2N$ degrees of freedom. It has a closed-form expression \cite{book_trees_1968_1}. For notation simplification, $\lambda(\alpha)$ will be defined as $\lambda$, the threshold scaling factor.

However, in a realistic radio telescope environment, the power $\sigma_n^2$ is not precisely known and that value fluctuates over time due to changes in the observational conditions (\textit{e.g.} antenna position, ionospheric conditions, frequency channel properties...). Thus, in order to maintain a constant $P_{fa}$, the threshold value must be continuously updated based on adaptive estimation of $\sigma_n^2$ and this estimation has to be robust against samples that belong to $H1$.

\citet{ravier_2005_1} and \citet{fridman_2008_1} have proposed several robust estimators of the power. Their performances are interesting but their implementations involve sorting and iterative loops which is not suitable in our 'on-the-fly' operational context. In the next subsection, a robust recursive power (RRP) estimator is presented.

\subsection{The robust recursive power estimator}

Let us define, $p(t)$, the instantaneous power of the signal under test (\textit{i.e.} $p(t)=\left| x_k(t) \right|^2$). Under the $H0$ hypothesis, the probability density function (pdf) of $p$ is given by:
\begin{equation}
\mbox{pdf}(p)= \frac{1}{\sigma_n^2} \exp\left(-\frac{p}{\sigma_n^2}\right) 
\label{eq:chi2}
\end{equation}
with $p \geq 0$.

If we just consider the power amplitudes, $\tilde{p}$,  below $\eta=\lambda.\sigma_n^2$, the clipped probability density function becomes:
 %and $\lambda$ is a scaling factor which defines the expected $P_{fa}$ according to Equ.~4x with N=1.
\begin{equation}
\mbox{pdf}(\tilde{p})= \frac{1}{\sigma_n^2 (1-\exp(-\lambda))} exp\left(-\frac{\tilde{p}}{\sigma_n^2}\right) 
\label{equ:chi2clip}
\end{equation}
with $0 \leq \tilde{p} \leq \lambda.\sigma_n^2$.

From, the previous equation, the mean, $\tilde{\sigma}_n^2$, of these clipped power amplitudes, $\tilde{p}$,  is easily derived:
\begin{equation}
\tilde{\sigma}_n^2=\sigma_n^2  \frac{1-\exp(-\lambda)-\lambda \exp(-\lambda)}{1-\exp(-\lambda)}
\label{eq:power_correction}
\end{equation}

Thus, the threshold $\eta$ can be redefined as:
\begin{equation}
\eta=\lambda.\underbrace{\frac{1-\exp(-\lambda)}{1-\exp(-\lambda)-\lambda \exp(-\lambda)}}_{\tilde{g}(\lambda)} \tilde{\sigma}_n^2
=\underbrace{\lambda.\tilde{g}(\lambda)}_{\tilde{\lambda}}\tilde{\sigma}_n^2
\label{eq:eta_tilde}
\end{equation}
where $\tilde{g}(\lambda)$ is the correction gain to apply to $\tilde{\sigma}_n^2$ to retrieve  $\sigma_n^2$ and $\tilde{\lambda}$ is the truncated threshold scaling factor. Table~\ref{tab:lambda} provides tabulated values of these parameters.
It is interesting to notice that:

\begin{eqnarray}
\lim_{\lambda \to 0} \tilde{\lambda}  =2 & \mbox{ and } & \lim_{\lambda \to 0} \tilde{\sigma}_n^2  =0 \\
\mbox{for } \lambda \gg 1 \mbox{, \bigskip    } \tilde{\lambda}  \simeq \lambda          & \mbox{ and } & \tilde{\sigma}_n^2   \simeq \sigma_n^2
\end{eqnarray}

\begin{table}[t]
 \centering
 \begin{tabular}{|c|c|c|c|c|}
  \hline
  $\lambda$ & $\tilde{g}(\lambda)$ & $\tilde{\lambda}$ & $\delta$ & \% clipped \\
  \hline
  \hline
   0.01 & 200.334 & 2.003 &  100.501 & 99.00 \\
   0.1  & 20.339 & 2.034 &  10.508 & 90.48\\
	1   & 2.392 & 2.392 & 1.582 & 36.79\\   
	1.5	& 1.757 & 2.635 &	1.287	&22.31 \\
	2	& 1.456 & 2.911 &	1.157 &	13.53 \\
	2.5	 & 1.288 & 3.220 &	1.089 &	8.21 \\
	3    & 1.187 &	3.560 &	 1.052 &	4.98 \\
  3.5 & 1.122 &	3.928 &	 1.031	& 3.02\\
  3.59352 & 1.113 &	4.000 & 	1.028 &	2.75 \\
  4	& 1.081 & 4.323 &	 1.019	& 1.83 \\
4.5 & 1.053 & 4.740 &	1.011 &	1.1 \\
5 & 1.035 & 	5.176 &	1.007 &	0.67 \\
7.9785	& 1.003 & 8 &	1.000 &	0.03 \\

  \hline
\end{tabular}
\caption{Tabulated values of  the threshold scaling factors $\lambda$, $\tilde{\lambda}$, the correction gain $\tilde{g}(\lambda)$, the freeze factor $\delta$ and the percentage of clipped samples under $H_0$ hypothesis.}
\label{tab:lambda}
 \end{table}

In this paper, we proposed to estimate  $\tilde{\sigma}_n^2$ by applying a low-pass Infinite Impulse Response (IIR) filter which is only activated when the power amplitude is below the running threshold $\eta(t)$ :

\begin{equation}
\tilde{\sigma}_n^2(t)= \left\{
 \begin{array}{ll}
        \beta.p(t)+(1-\beta). \tilde{\sigma}_n^2(t-1) & \mbox{if }  p(t)<\underbrace{\tilde{\lambda}. \tilde{\sigma}_n^2(t-1)}_{= \eta(t)}  \\
       \tilde{\sigma}_n^2(t-1) & \mbox{else.}
    \end{array}
    \right.
    \label{equ:clip_power_est}
\end{equation}

where $ \tilde{\sigma}_n^2(t)$ is the robust estimation of $\tilde{\sigma}_n^2$ at time $t$, $\beta$ ($0<\beta<1$) is the forgetting factor which drives the time response of the estimator.  
This estimator must be initialized with $ \tilde{\sigma}_n^2(0) \geq \tilde{\sigma}_n^2$ (\textit{e.g.} $\max(\left| x_k(t) \right|^2)$ or the maximum value used for computation).

The following property is easily checked under stationary hypothesis:
\begin{equation}
\lim_{t \to \infty} \tilde{\sigma}_n^2(t)  =  \tilde{\sigma}_n^2
\end{equation}

Figure~\ref{fig:compar_IIR_FIR} provides an illustration of the previous equations. It shows also the robustness against RFI. Under $H1$ hypothesis, this recursive variance estimator will mitigate the outliers influence and provide a robust  estimation of the expected power under the $H0$ hypothesis. In the next sections, some operational properties of our IIR filter are derived.

\subsubsection{The equivalent estimation window length}

Instead of the IIR defined in Eq.~\ref{equ:clip_power_est}, a simple low pass Finite Impulse Response (FIR) filter could have been used:

\begin{equation}
\hat{\sigma}_n^2(t)=\frac{1}{N}\sum_{k=0}^{N-1} \tilde{p}(t-k)
\label{eq:FIR}
\end{equation}

Since its implementation needs more memory resources, the IIR implementation has been preferred but it would be interesting to evaluate an equivalent window length $N$  for the IIR approach as a function of the forgetting factor $\beta$.

For this purpose, the variance of both approaches will be compared, assuming first that $p(t)$ is replaced by $\tilde{p}(t)$ in Eq.~\ref{equ:clip_power_est} (\textit{i.e.} clipped samples are discarded).

Thus,  the variance of $\tilde{\sigma}_n^2(t)$, $var(\tilde{\sigma}_n^2)$ is given by
\begin{equation}
 var(\tilde{\sigma}_n^2)= \frac{\beta}{2-\beta}.var(\tilde{p})
\label{eq:FIR_var}
\end{equation}

with $var(\tilde{p})=\left( 1-\frac{\lambda^2 \exp(-\lambda)}{(1-\exp(-\lambda))^2} \right).\sigma_n^4$

\begin{figure}[t]
\begin{center}
\includegraphics[width=\linewidth]{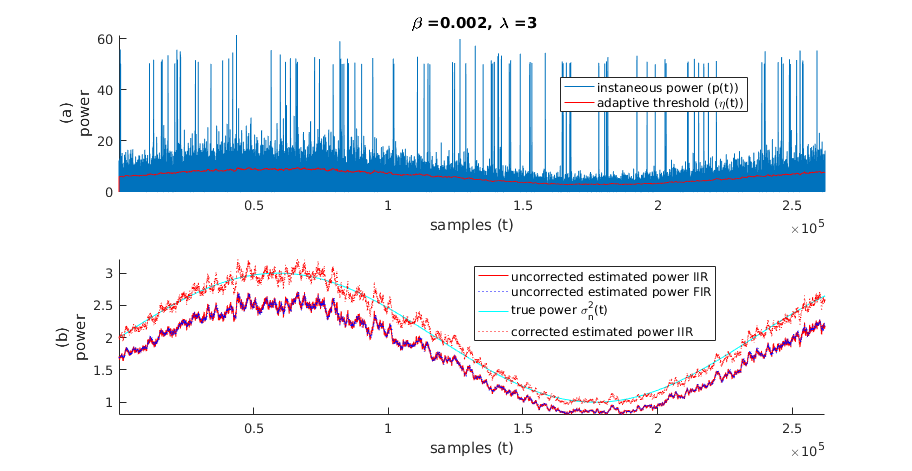} %100 percent
\end{center}
\caption{Simulation of the robust recursive power (RRP) estimator under impulsive RFI pollution and non stationary hypothesis (\textit{i.e.} the true noise power presents a sine wave fluctuation). \textbf{(a)} The corresponding instantaneous power, $p(t)$, (blue line) and the running threshold, $\eta(t)$, (red line) computed with eq.~\ref{equ:clip_power_est}. The estimation parameters are $\beta = 0.002$ and $\lambda=3$. The impulsive RFI are clearly visible but have no impact on the threshold estimation. \textbf{(b)} The RRP estimator (red line), $\tilde{\sigma}^2_n(t)$, is compared with the FIR one (dashed blue line),  $\hat{\sigma}^2_n(t)$, defined by Eq.~\ref{eq:FIR}. The perfect matching is obtained through Eq.~\ref{eq:N}. The amplitude shift compared to  the true noise power, $\sigma^2_n(t)$, (cyan line) is related to the clipped samples. By applying the correction defined in Eq.~\ref{eq:power_correction}, the corrected RR estimator (dashed red line) matches the true noise power.}
\label{fig:compar_IIR_FIR}
\end{figure}

By matching this variance with the one obtained from Eq.~\ref{eq:FIR}, the equivalent estimation window length is defined by:

\begin{equation}
 N=\frac{2}{\beta}-1
 \label{eq:N}
\end{equation}

which becomes $N=\frac{2}{\beta}$ when $\beta << 1$.  Figure~\ref{fig:compar_IIR_FIR}.b provides an illustration of  this result. 

To get the true equivalent estimation window length, $\tilde{N}$, we have just to take into account the  number of samples which freeze the estimator. In other words, $N$ has to be multiplied by the freeze factor:
\begin{equation}
 \delta=\frac{1}{1-\exp(-\lambda)}
 \label{eq:freeze_factor}
\end{equation}

For example, let us consider a SOI stationary time interval of 1~s.  Assuming 200~kHz bandwidth  for the beamlet as in the EMBRACE case, the reference power should be estimated over $\tilde{N}=2. 10^5$ samples, which gives $N=190040$ non-clipped samples with $\lambda=3$ according to Table~\ref{tab:lambda} and finally $\beta=1.052 \mbox{ } 10^{-5}$ for the robust estimator defined by Equ.~\ref{equ:clip_power_est}. %Lowered $\lambda$ will reduce the number of available samples and consequently  will increase the estimator variance.   Choosing $\lambda$ value lower than 1 will induce 50\% extra samples to converge.

\subsubsection{The synchronisation delay}

Due to the IIR filter latency, the input data $p(t)$  has to be resynchronized with the RRP estimator. 

 Assuming slow variations of the reference power and by taking into account, $\delta$, the freeze factor, the filter latency, $\tilde{\tau}$, can be derived from the phase response of the IIR filter (see appendix):
 \begin{equation}
\tilde{\tau}=\frac{\frac{1}{\beta}-1}{1-\exp(-\lambda)}
\label{eq:tau_tilde}
 \end{equation}

\subsection{The Bernoulli power detector}
As already mentioned, real-time requirements constrain the complexity of  RFI detection schemes.  Constraints include : limitation of resources usage, data flow continuity, and data synchronization continuity. 
%The following subsections describe different approaches which, all, assume that the RRV estimator provides the reference power expected under the $H_0$ hypothesis.

\begin{figure}[t]
\begin{center}
\includegraphics[width=0.75\linewidth]{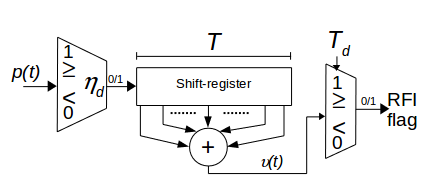} %100 percent
\end{center}
\caption{Schematic view of the Bernoulli power detector. $p(t)$ is the instantaneous power. $\eta_d$ is the RFI power threshold. Depending on $\tilde{\sigma}_n^2(t)$. The 1-bit shift-register is $T$ samples long. The RFI flag is set to 1 when the number of 1 in the shift-register ($\nu(t)$) is greater or  equal to $T_d$.}
\label{fig:power_detector}
\end{figure}

In this context, we propose the following approach (see Figure~\ref{fig:power_detector}): a given window made of $T$ samples $p(t)$ is flagged as polluted when the number of outliers, defined as $p(t)\geq \eta_d$, exceeds a threshold $T_d$. The strategy is to implement large power threshold  $\eta_d$, short time window $T$ and to choose {$T_d \approx T$} for strong RFI, but to implement smaller  power threshold $\eta_d$, longer time window $T$ and to choose {$T_d<T$} for weak RFI. Of course, any intermediate configuration can be set up as well. All these  detectors will run simultaneously and individual flags will be merged into a final flag after resynchronisation.  

Let us define $p_{\eta_d}$ the probability to have $p(t)\geq \eta_d$. With Eq.~\ref{eq:chi2}, we get $p_{\eta_d}=\exp(-\frac{\eta_d}{\sigma_n^2})$ which gives $p_{\eta_d}=\exp(-\lambda_d)$ where $\lambda_d$ follows Eq.~\ref{eq:eta_th}. 
Then, the probability to have at least $T_d$ samples over $T$  above $\eta_d$ is a Bernoulli distribution given by:
\begin{equation}
\mbox{Proba}(\nu(t) \geq T_d)=\sum^T_{k=T_d} \frac{T!}{k!(T-k)!}p_{\eta_d}^k (1-p_{\eta_d})^{T-k}=I_{p_{\eta_d}}(T_d,T-T_d+1)
\label{eq:pfa_bernoulli}
\end{equation}
where $\nu(t)$ is the sum of the binary flags shift-register content, $I_{p_{\eta_d}}(a,b)=\frac{B(p_{\eta_d};a,b)}{B(a,b)}$,  $B(a,b)$ is the beta function, and $B(x;a,b)$ is the incomplete beta function. 

With this equation, we can compute the expected detector false alarm probability according to the parameters $\lambda_d$, $T$  and  $T_d$. In practice, we will have access to $\tilde{\sigma}_n^2$ via the RRP estimator, thus the truncated  threshold scaling factor $\tilde{\lambda}_d$ will be preferably implemented. In the same way as for  Eq.\ref{eq:eta_tilde}, we get:
\begin{equation}
\tilde{\lambda}_d=\tilde{g}(\lambda).\lambda_d
\end{equation} 
where $\lambda$ is the scaling threshold factor used for the RRP estimator.  Table~\ref{tab:lambda_detector} provides tabulated values of these parameters for $\lambda=3.59352$.

For the following case, $T=T_d$, the corresponding false alarm probability simplifies to $P_{fa}=\exp \left(-\lambda_d T \right)$. 

\begin{table}[t]
 \centering
 \begin{tabular}{|c|c|c|}
  \hline
  $\lambda_d$  & $\tilde{\lambda}_d$ & $p_{\eta_d}$  \\
  \hline
  \hline
   0.01 & 0.011  & 99.00 \\
   0.1  & 0.111&  90.48\\
   \textbf{0.81415} & $\mathbf{\frac{29}{32}}$ & \textbf{44.30} \\
	1   & 1.113 & 36.79\\   
	1.5	& 1.670 	& 22.31 \\
	2	& 2.226  &	13.53 \\
	2.5	 & 2.783 &	8.21 \\
	3    & 3.339 &	4.98 \\
  3.5 & 3.896	& 3.02\\
  \textbf{3.59352} & \textbf{4} &	\textbf{2.75} \\
  4	& 4.452	& 1.83 \\
4.5 & 5.009 &	1.1 \\
5 & 5.566 &	0.67 \\
7.187	& 8 &	0.03 \\

  \hline
\end{tabular}
\caption{Tabulated values of  the threshold scaling factors $\lambda_d$, $\tilde{\lambda}_d$,  and $p_{\eta_d}$,  the percentage of false alarm under $H_0$ hypothesis. The RRP estimator has been set up with $\tilde{\lambda}=4$ so that $\tilde{g}=1.113$ (see Table~\ref{tab:lambda}).}
\label{tab:lambda_detector}
 \end{table}

A simulation with two Bernoulli power detectors has been set up using the following parameters:
\begin{itemize}
\item \textbf{RRP estimator:} $\tilde{\lambda}=4$ and $\beta=2^{-11}$.
\item \textbf{weak Bernoulli power detector:} $\tilde{\lambda}_d=\frac{29}{32}$, $T=30$ samples  and  $T_d=25$ samples.
\item \textbf{strong Bernoulli power detector:} $\tilde{\lambda}_d=4$ , $T=3$ samples  and  $T_d=3$ samples.
\end{itemize}

\begin{figure}[h]
\begin{center}
\includegraphics[width=\linewidth]{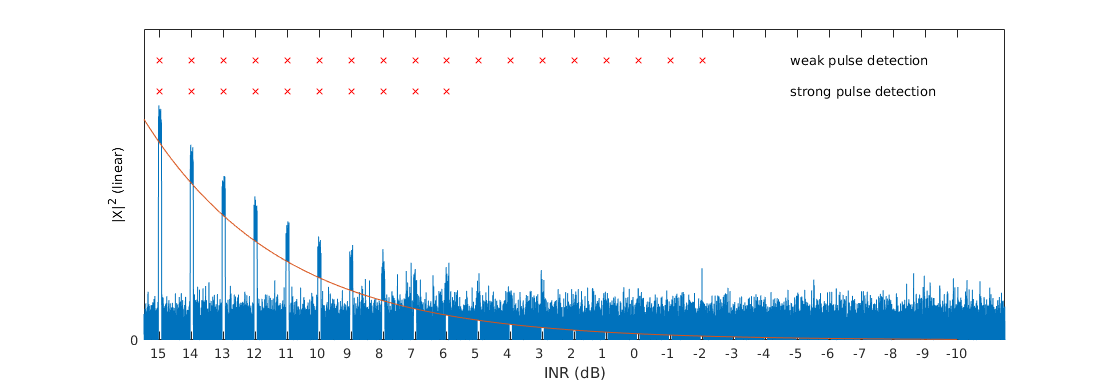} %100 percent
\end{center}
\caption{Simulation of the Bernoulli power detector. The RRP estimator (parameters $\tilde{\lambda}=4$ and $\beta=2^{-11}$) associated with two Bernoulli power detectors (parameters $\tilde{\lambda}_d$, $T$  and  $T_d$ equal to $[4, 3, 3]$ for the strong detector and $[\frac{29}{32}, 30, 25]$ for the weak detector respectively).  The red ''x'' indicate only the \emph{true positives}. The \emph{false positives} are not plotted (see Table~\ref{tab:pfa_simu}).}
\label{fig:detetect_vs_INR}
\end{figure}

A white noise affected by intermittent RFI pulses with decreasing INR has been generated. Figure~\ref{fig:detetect_vs_INR} shows the detection performance with the previous parameters. The strong pulse detector and the weak pulse one succeed in detecting pulses with INR values lower than 6~dB and -2~dB respectively. A comparison between the detector probability of false alarm in theory and evaluated after simulation is provided in Table~\ref{tab:pfa_simu}.

\begin{table}[h]
 \centering
 \begin{tabular}{|c|c|c|}
  \hline
  Detector type &  $p_{fa}$ (Theory) & $p_{fa}$ (Simulation) \\
  \hline
  \hline
  strong detector & $2.1 \mbox{ x } 10^{-3}$ \% &  $2.0963 \mbox{ x } 10^{-3}$ \% \\
  \hline
  weak detector & $1.3 \mbox{ x } 10^{-3}$ \% &  $1.3962 \mbox{ x } 10^{-3}$ \% \\
  \hline
\end{tabular}
\caption{False positive comparison between theory and simulation. The simulation parameters are the same as the one in Figure~\ref{fig:detetect_vs_INR}. Theoretical $p_{fa}$ are given by Eq.~\ref{eq:pfa_bernoulli} where $p_{\eta_d}$ is extracted from Table~\ref{tab:lambda_detector}.}
\label{tab:pfa_simu}
 \end{table}

\subsection{First real time experiment}

An 1-channel implementation (\textit{i.e.} K=1) of this detection scheme has been conducted at the Nan\c{c}ay Observatory \cite{Ait-allal_2010_1}. It was designed for real time radar blanking at the Nan\c{c}ay Decimeter Radio Telescope (NRT). Radar pulses are often an issue in radio astronomical observations. \citep{art_niamsuwan_2005_1} has proposed some simple pulse detectors but with basic performances and \citep{art_dong_2005_1} has defined a much more advanced algorithm that gives better performances for its finest setting. However, it includes a lot of information about the specific radar pulse shape. Any radar pulse whose shape differs significantly from the model will not be detected as easily. In our approach, radar pulse time characteristics will be used to define  $T$  and  $T_d$.

Thus, with the same parameters as the previous simulation, the RRP estimator associated with two Bernoulli power detectors has been implemented into  a Virtex-II \textregistered{Xilinx} FPGA.  Less than 3\% of the FPGA resources were used. The resulting RFI flag was used to blank 2048-bins Fast Fourier Transform blocks in the final spectrometer integration leading to a final $pfa$ of 4.0\% and 6.4\% for the weak and strong pulse detector, respectively.

\begin{figure}[t]
\begin{center}
\includegraphics[width=0.75\linewidth]{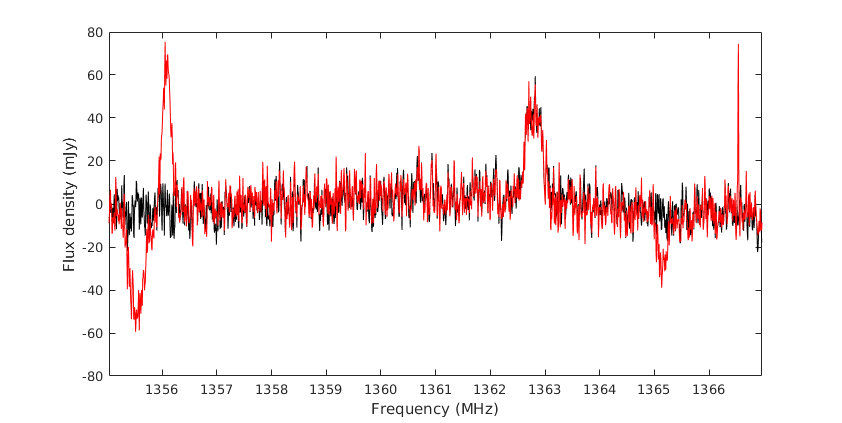} %100 percent
\end{center}
\caption{Test of the radar blanker with real data and in real time. The observed source is PGC51094. The red curve
is the spectrum obtained without blanking and the black one is the spectrum obtained with the blanker active.}
\label{fig:PGC51094}
\end{figure}

Figure~\ref{fig:PGC51094} illustrates the effectiveness of this approach on real observations in the presence of radar interference. The observed bandwidth was 14~MHz wide.

In the next section, we present the ongoing implementation of all these detectors and the RRP estimator in the Uniboard. They are used to continuously monitor the data quality at different stages of a 64-bins spectrometer analysing 248 complex 200~KHz beamlets generated by the EMBRACE radio telescope. The objective is to test both  the relevance of an integrated RFI database and the concept of an oriented accumulation.

\section{EMBRACE and UniBoard Experiment} \label{firmware}

\subsection{Overview}
EMBRACE is a phased-array radio telescope \citep{torchinsky_2016_1}. Based on a LOFAR station firmware, it provides a set of 200~kHz digital single polarization waveforms from 2 independent beams. In our application, 248 waveforms are used as data input for Uniboard. Besides, EMBRACE by itself provides some regular statistics as well (100~MHz bandwidth power statistics for all analog inputs and 200 kHz correlation statistics) on the received data. These statistics will be added in the RFI monitoring system as well, but it is outside of the scope of this paper.

In radio astronomy, the elementary time-frequency slot shape is driven by radio telescope design considerations, basically from wide bandwidth in the front end to narrow bandwidth in the backend (see Figure~\ref{fig:overview}). In our experiment, we have set up a similar scheme but at a smaller scale. 

\begin{figure}[t]
\begin{center}
\includegraphics[width=\linewidth]{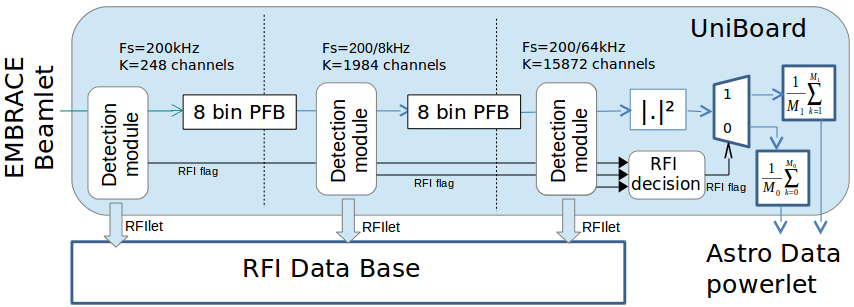} %100 percent
\end{center}
\caption{Overview of the UniBoard firmware. The detection modules are described in Figure~\ref{fig:overview_detection}. Fs is the sampling frequency at each stage. Each polyphase filter bank (PFB) splits the input channel into 8 sub-channels. These sub-channels are decimated by 8. At each stage, detectors analyse the waveforms and deliver RFI information to the RFI database and toward the RFI decision module. The latter will control the final spectral accumulation. The accumulation number, $M$, is constant ($M=M_0+M_1$).}
\label{fig:overview_uniboard}
\end{figure}

Thus, UniBoard has been configured for two main tasks (see Figure~\ref{fig:overview_uniboard}). The first one is the data quality monitoring of these beamlets at different time-frequency resolutions (200~kHz, 200/8~kHz and 200/64~kHz). For this purpose, each beamlet is decomposed into 64 channels with two successive 8-bin polyphase filter banks (PFB). These PFB are maximally decimated (\textit{i.e.} the output rate is the input rate divided by 8). At each stage, a detection module monitors the data and provides RFI information about data quality.  This information, named RFIlet, is sent to the RFI database.
The proposed multi-resolution approach is a way to improve the scaling between detector time-frequency resolution and RFI characteristics. For example, assuming white noise and a narrow band interferer, the interference to noise ratio (INR) will be 18 dB better at the output than at the input.

The second one consists in delivering of 64-bins power spectral estimators for all the 200~kHz input beamlets. The spectrum accumulation is driven in real time by the previous data quality analysis. This approach is called \emph{oriented accumulation}. Clean data and flagged data are accumulated separately. In the current implementation, the RFI flagging is binary but a multilevel decision could be easily implemented as well. Consequently, no data are lost and the end-user has the option to use or not the flagged data. The RFI decision process is based on inheritance: once a sample from a given channel is detected as polluted, all sub-channels linked with this sample are flagged as well. It is a rather conservative approach but according to the RFI context other rules could be implemented.

\subsection{UniBoard Detection module firmware}

\begin{figure}[t]
\begin{center}
\includegraphics[width=\linewidth]{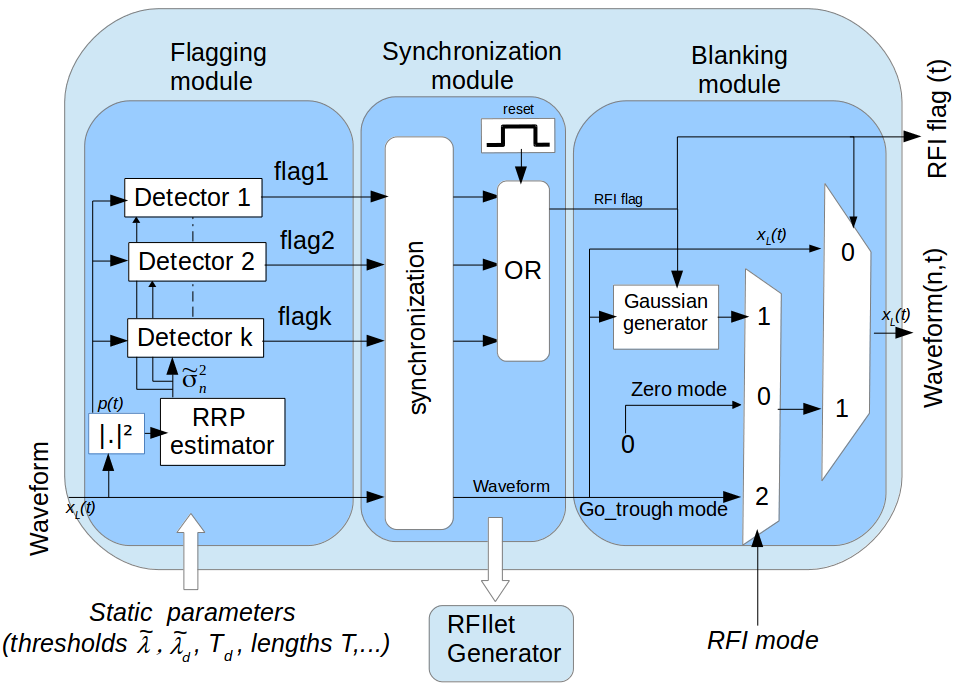} %100 percent
\end{center}
\caption{Overview of the detection module. Each detector  evaluates  the RFI  contamination level. Then, the corresponding flags and the waveforms are sent to the synchronized decision module. The synchronisation  removes the differential latencies generated by detector time responses.  All detector flags are merged into one global RFI decision. }
\label{fig:overview_detection}
\end{figure}

\begin{figure}[t]
\begin{center}
\includegraphics[width=\linewidth]{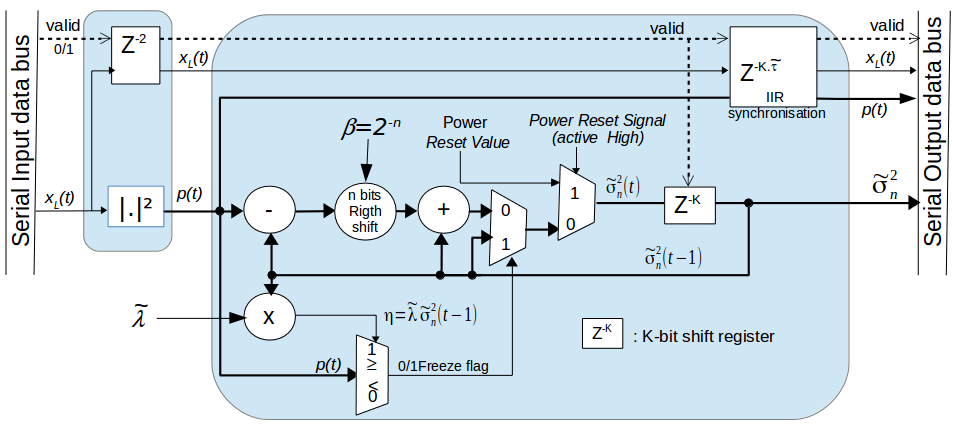} %100 percent
\end{center}
\caption{Overview of the RRP estimator module. $K$ is the channel number at the considered stage, $\tilde{\lambda}$ is the threshold scaling factor defined in Eq.~\ref{eq:eta_tilde}, $\beta$ is the IIR forgetting factor and $\tilde{\tau}$ is the synchronization delay defined by Eq.\ref{eq:tau_tilde}.  The valid flag is attached to each serial data bus slot and it validates the presence of valid $x_k(t)$ data.}
\label{fig:RRP_implementation}
\end{figure}

\begin{figure}[t]
\begin{center}
\includegraphics[width=\linewidth]{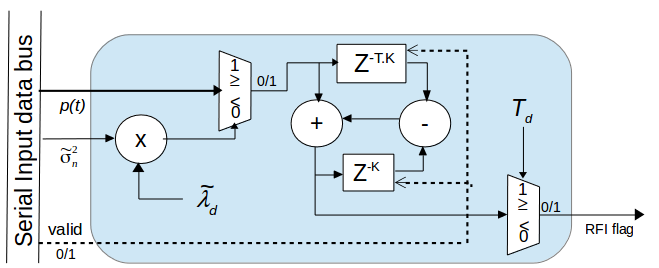} %100 percent
\end{center}
\caption{Overview of the Bernoulli power detector module. $K$ is the channel number at the considered stage, $\tilde{\lambda}_d$ is the threshold scaling factor defined in Eq.~\ref{eq:eta_tilde}, $T$ is the detection window length. The RFI flag is set to 1 when the number of individual flagged power samples is greater or  equal to $T_d$. The valid flag is attached to the input data, $x_k(t)$, through the serial data bus and it validates the presence of valid $p(t)$ data}
\label{fig:power_detector_implementation}
\end{figure}

Figure~\ref{fig:overview_detection} gives an overview of an RFI detection module. It includes: 
\begin{itemize}
\item \textbf{a flagging module:} each detector analyzes a set of $T$ consecutive time samples from a given channel and triggers a flag if the threshold is reached. The analysis window lengths, $T$, and the thresholds, $T_d$ and $\tilde{\lambda}_d$, are static parameters. In the current implementation, the detectors are based on the Bernoulli power detector scheme (see Figure~\ref{fig:power_detector_implementation}). All these detectors are calibrated by the RRP estimator (see Figure~\ref{fig:RRP_implementation}).

\item \textbf{a synchronization module:} Each detector generates its own latency (mainly parametrized by $T$). Consequently, all RFI flags and waveforms must be synchronized with the longest latency. This is one of the tricky parts of the design and it is very technical. Therefore, we do not provide a detailed analysis of this module. The RFI flags provided by the Bernoulli power detector modules flag only the last sample of their respective detection windows. The purpose of this synchronization module is to expand the flags all over the detection windows and then to merge all these signals into one final RFI flag. A simple OR function is used in this version but more complex rules could be implemented. This module is connected with the RFIlet generator which communicates with the RFI database. %The reset duration is based on Eq.~\ref{eq:reset}.

\item \textbf{A blanking module:}  3 modes are available (part of the static configuration):
\begin{itemize}
  \item \textbf{Go-through mode:} waveform time samples are flagged, the waveforms are not modified.
  \item \textbf{Zero mode:} flagged waveform time samples  are replaced by zeros
  \item \textbf{Gaussian mode:}  flagged waveform time samples are replaced by synthesized complex Gaussian samples of the same variance (not implemented in the current firmware version).
\end{itemize}
\end{itemize}

For the Uniboard RRP estimator implementation, Eq.~\ref{equ:clip_power_est} has been reordered in the following way:
\begin{equation}
\tilde{\sigma}_n^2(t)= \left\{
 \begin{array}{ll}
        \beta. \left( p(t)-  \tilde{\sigma}_n^2(t-1)  \right) +  \tilde{\sigma}_n^2(t-1) & \mbox{if }  p(t)< \tilde{\lambda}. \tilde{\sigma}_n^2(t-1)  \\
       \tilde{\sigma}_n^2(t-1) & \mbox{else.}
    \end{array}
    \right.
    \label{equ:clip_power_est_modif}
\end{equation}
where $\beta=2^{-n}$  so that the corresponding multiplier can be implemented as a simple bit-shift with no resource cost. Thus, for $n \gg 1$, the equivalent estimation window length, $N \simeq 2^{n+1}$. $\beta$ is a static parameter.

With the same considerations, it could be convenient to choose the static threshold scaling factor $\tilde{\lambda}$ as a power of 2. According to Table~\ref{tab:lambda}, two values are valid, $\tilde{\lambda}=4$ or $\tilde{\lambda}=8$, which gives a percentage of clipped samples equal to 2.75\% and 0.03\%, respectively.  

Figure~\ref{fig:RRP_implementation} shows the detailed design. A valid signal is attached to the waveform data and is used to freeze the design when no valid data is available at the input. All data are time stamped to keep track of the data stream integrity. All this information is put into a serial data bus which is compliant with the Altera Avalon standard \cite{altera_2015}. To speed up the design, it is possible to insert registers between the operators. This will modify the IIR filter response but assuming slow variations of $\tilde{\sigma}_n^2$, all the previous study will be still valid. With similar considerations, it is also possible to remove the IIR resynchronization module. This module is based on Eq.~\ref{eq:tau_tilde}.

The synthesis report indicates a maximum clock frequency of 250~MHz (we need 180~MHz). The amount of multipliers is low, especially if power-of-2 scaling factors are set up. The memory requirement is high since it depends on $K$, the number of channels. However it is manageable especially if we consider  slow variations of $\tilde{\sigma}_n^2$ so that we can remove the IIR resynchronization module.

\section{Conclusions and perpectives} \label{conclusion}

SKA will be built in a quiet zone far from inhabited areas, except perhaps for some remote stations. SKA will have to face intermittent interference from satellite systems and airplanes. These signals are relatively strong.

Our approach is based on multi-scale elementary time-frequency slot shapes driven by radio telescope design considerations. Furthermore, for optimal RFI detection performances, time-frequency slots should be adapted to RFI time-frequency shapes. In this case, the INR is maximized and all the RFI time-frequency characteristics, which can be used for detection, are preserved. We have proposed a low-cost power detection scheme which can be implemented in the pre-correlation stages at different levels from large bandwidth in the front-end to narrow bandwidth in the backend. This RFI monitoring system can be used to feed an RFI database and/or trigger for other more complex RFI mitigation techniques. Thus, the digital signal processing resources could be fully dedicated to regular signal processing tasks most of the time and could be partially re-used (scheduled) for observations facing specific RFI issues and requiring specific RFI mitigation techniques.

A first FPGA implementation designed for a specific time-frequency resolution  level has proved the effectiveness of the proposed power detection scheme. Our RRP estimator provides a robust and adaptive estimation of the reference power even in presence of intermittent RFI, and our Bernoulli power detector can achieve low INR RFI detection at a very low implementation cost.

A second implementation based on a high performance computing board, UniBoard, has been described. The objective is to generate a multi-level detection scenario based on data received by the EMBRACE radio telescope. At each level, our power detection scheme is implemented and the resulting RFI flags are used to drive the final spectrum integration and to feed an RFI database. This experiment is still in progress.

\begin{figure}[h]
\begin{center}
\includegraphics[width=0.5\linewidth]{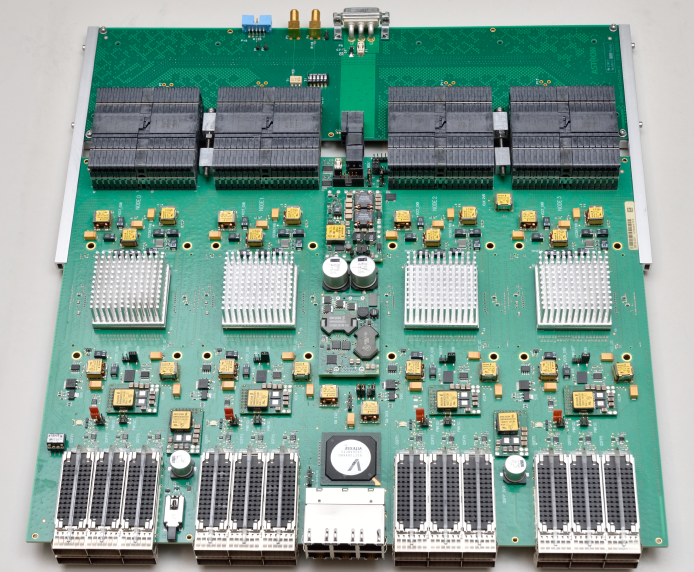} %100 percent
\end{center}
\caption{The high-performance computing platform, UniBoard$^2$, for future radio-astronomical instruments developed under a Joint Research Activity in the RadioNet3 FP7  European programme. It is currently based on 4 \textregistered{Altera} Arria 10 but it can be upgraded to Stratix 10 providing up to 28~TMACS per board. 2~Tbps into front side generic interfaces (1, 10 and 40G Ethernet, DDR4 memory). The board implements also 4 Hybrid Memory Cubes (HMC) in a mesh on an HMC Extension Module.}
\label{fig:uniboard2}
\end{figure}

The power based detectors are widely implemented in practice since no RFI a priori information is used. However, to improve the detection performances, we plan to implement more  specific detectors based on ad-hoc RFI characteristics such as non Gaussianity \cite{art_nita_2007_1}, spectral properties \cite{art_hellbourg_2017_1}, cyclostationary properties \cite{art_weber_2007_1}. 
As these implementations will used more resources, a new generation UniBoard will be targeted. This Uniboard2 (see Figure~\ref{fig:uniboard2}), also developed under a Joint Research Activity in the RadioNet3 FP7  European programme, is based on 4 \textregistered{Altera} Arria 10. It provides up to 3.6~TMACs and 2~Tbps into front side generic interfaces.

\section*{Acknowledgments}
The authors would like to thank the European Commission Framework Program 7 ( Project Radionet Uniboard, contract no 227290 and project Radionet Uniboard2, contract no 283393) for funding this research on RFI mitigation and its hardware implementation.

\section*{Appendix : The synchronisation delay} \label{app:synchro}

The transfer function of the IIR filter defined in Eq.~\ref{equ:clip_power_est} is given by:
\begin{equation}
H(z)=\frac{\beta}{1-(1-\beta).z^{-1}}
\end{equation} 
Thus, the frequency response is given by:
\begin{equation}
H(z)|_{z=\exp(\jmath 2 \pi f)}=H(f)=\frac{\beta \left[ 1 - (1-\beta) \cos (2 \pi f)-\jmath (1-\beta)\sin (2 \pi f)\right]}{\left[ 1 - (1-\beta) \cos (2 \pi f)\right]^2 + (1-\beta)^2 \sin^2((2 \pi f) }
\end{equation}
From the previous equation, the phase response can be derived :
\begin{equation}
\phi(f)=\arctan \left( \frac{(\beta-1)\sin (2 \pi f)}{ 1 + (\beta-1) \cos (2 \pi f)} \right)
\end{equation}
 Assuming slow variations of the reference power (\textit{i.e.} $f \simeq 0$), we get the following approximation
 \begin{equation}
\phi(f)=2 \pi f \left(1- \frac{1 }{\beta} \right)
\end{equation}
From the previous equation, the response time delay can be derived:
\begin{equation}
\tau=-\frac{1}{2 \pi} \frac{\partial \phi (f)}{\partial f}=\frac{1}{\beta}-1
\end{equation}
By taking into account the frozen samples (see Eq.~\ref{eq:freeze_factor}), the true delay is given by:
\begin{equation}
\tilde{\tau}=\frac{\frac{1}{\beta}-1}{1-\exp(-\lambda)}
\end{equation}

\bibliographystyle{ws-jai}

\bibliography{weber}

\end{document}